\documentclass[a4paper]{article}

\usepackage{icrc2013}

\usepackage{graphicx}
\usepackage[english]{babel}
\usepackage{hyperref}

\newcommand{\onees}{\textit{1ES~0229+200}}
\newcommand{\fermi}{\textit{Fermi}}
\newcommand{\fermilat}{\textit{Fermi}-LAT}
\newcommand{\lat}{\textit{LAT}}
\newcommand{\swift}{\textit{Swift}}
\newcommand{\swiftxrt}{\textit{Swift-XRT}}
\newcommand{\swiftbat}{\textit{Swift-BAT}}
\newcommand{\swiftuvot}{\textit{Swift-UVOT}}
\newcommand{\uvot}{\textit{UVOT}}
\newcommand{\xrt}{\textit{XRT}}
\newcommand{\bat}{\textit{BAT}}
\newcommand{\veritas}{VERITAS}
\newcommand{\hess}{\textit{H.E.S.S.}}
\newcommand{\rxte}{\textit{RXTE}}

\newcommand{\aap}{A\&A}
\newcommand{\aaps}{A\&AS}
\newcommand{\apj}{ApJ }
\newcommand{\apjl}{ApJL}
\newcommand{\mnras}{MNRAS}

\title{VERITAS results from a three-year observing campaign\\ on the BL Lac object 1ES 0229+200}

\shorttitle{VERITAS results on 1ES0229+200}

\authors{
Matteo Cerruti $^{1}$,
for the VERITAS Collaboration.
}

\afiliations{
$^1$ Harvard-Smithsonian Center for Astrophysics; 60 Garden Street, Cambridge, MA 02138, USA 
}

\email{matteo.cerruti@cfa.harvard.edu}

\abstract{The BL Lac object \onees\  is a firmly established very-high-energy (VHE; E $>$ 100 GeV) source. Located at a redshift of 0.1396, its published VHE spectrum is particularly hard ($\Gamma \simeq 2.5$) and is widely used to perform studies on the extragalactic background light (EBL) and the intergalactic magnetic field (IGMF). We present here the results of a long-term observing campaign on \onees\ with \veritas, lasting from 2009 to 2012, supported by multi-wavelength observations. We discuss the \veritas\ results with a particular emphasis on the implications for the synchrotron-self-Compton modeling of the source emission and for the theoretical studies (especially on the intergalactic magnetic field) based on its VHE spectrum.}

\keywords{Gamma-rays: observations; Galaxies: active; Blazars; 1ES 0229+200}

\begin{document}
\maketitle

\section{Introduction}
The detection of the hard-spectrum ($\Gamma \simeq 2.5$), distant ($z=0.1396$) blazar \onees\ at very high energies (VHE; E $>$100 GeV) by \hess\ in 2007 \cite{Aharonian07} generated excitement among the members of the VHE community. It was well known that VHE $\gamma$-rays are attenuated via pair production on the EBL as they propagate through the Universe (see \cite{Gould67}) and that the pairs are then deflected by the IGMF (see \cite{Neronov09}). However, the majority of the models of the EBL at the time postulated a strong EBL and a relatively nearby $\gamma$-ray horizon (see for example \cite{Stecker06}). The discovery of a distant blazar with a hard TeV spectrum favored low-EBL models, close to the lower limits obtained by galaxy counts. \\
Distant, hard-spectrum blazars are also ideal for studies of the IGMF for similar reasons. The pairs produced in EBL interactions are deflected by the IGMF before interacting with cosmic microwave background (CMB) photons via inverse-Compton scattering. If the IGMF is not overly strong, the resulting high-energy and VHE photons are directed along the path of the original emitted photon (for a discussion of this, see \cite{Arlen}). This effect can cause a delay in the arrival of the signal and extended emission around point sources (for a review of these processes see \cite{Neronov09}). The effect of the IGMF on VHE spectra is difficult to disentangle from the features seen due to the EBL, since the IGMF will reprocess TeV photons into the GeV range, which will attenuate and soften the emitted VHE spectrum. Significant effort has been made to place limits on the IGMF using VHE and high-energy blazars by comparing the flux seen in the two energy bands (for example \cite{Chuck11}). Since the reprocessing occurs over time (the exact time depends on the IGMF strength, coherence length and distance to the source), these arguments usually depend on the VHE flux not varying, at least during the period of observation.\\
The lack of historical evidence of VHE variability was used by some authors to justify using measurements of \onees\ to constrain the strength of the IGMF (\cite{Arlen, Chuck11, Huan11, Neronov10}).\\ 
We present here the results from a long-term VHE study over three seasons of this unique blazar using \veritas, as well as multi-wavelength observations in visible light, UV and X-rays. We present a standard synchrotron-self-Compton (SSC) modeling of the source emission, and discuss the repercussions of the \veritas\ measurement on EBL and IGMF studies.\\

\section{\veritas\ observations}
\veritas\ is a ground-based imaging atmospheric Chernekov telescope (IACT) array located at the Fred Lawrence Whipple Observatory (FLWO) in southern Arizona (31 40N, 110 57W,  1.3km a.s.l.). For more details about the instrument, see \cite{Holder11}. \\

The \veritas\ collaboration has initiated a long-term plan which includes the observation of relatively distant blazars with hard spectra. The goal of this strategy is to build up a database of spectral energy distributions (SEDs) from a variety of blazars whose emission can carry the signature of the EBL it traverses and to study the blazar population in greater detail. As part of this program, \veritas\ observed \onees\ for a total time of 54.3 hours. These observations were taken over three seasons (2009 to 2012) and resulted in a strong detection of 11.7 $\sigma$.  VERITAS observed this source in a wobble configuration, where the telescopes are pointed 0.5 degrees away from the source so that a simultaneous background sample can be taken along with the on-source observations \cite{Fomin} (the field of view of \veritas\ being 3.5$^\circ$).\\

\begin{figure}[t!]
\vspace{0.3cm}
\includegraphics[width=240pt]{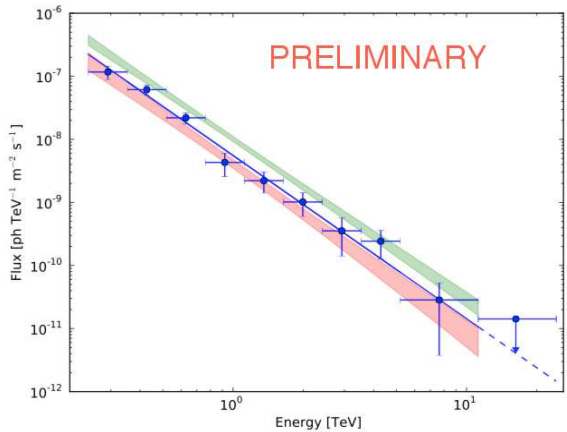}
\caption{The measured VHE spectrum from \onees\ averaged over all three seasons (blue points with error bars). The upper (green) and lower (red) shaded regions show the spectral shape during the high (MJD 55118 - 55131) and low (MJD 55144 - 55951) periods, respectively (see Fig. \ref{VERITASLC})} \label{VERITASspec}
\end{figure}

\begin{figure}[t!]
\vspace{0.3cm}
\includegraphics[width=240pt]{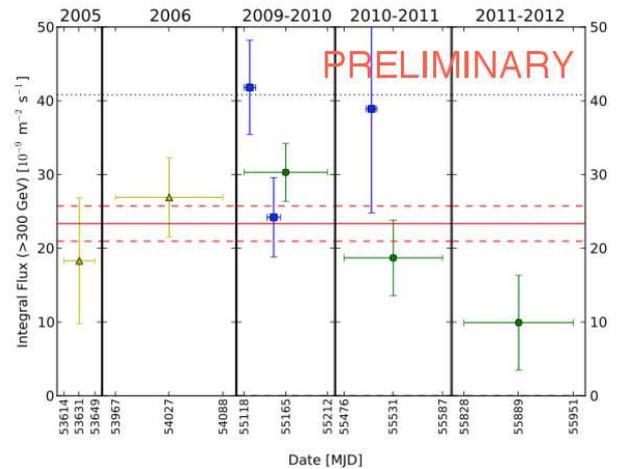}
\caption{ Integral flux above 300 GeV for \onees\, binned by observing season (green circles). The yellow triangles in 2005 and 2006 are from the previous \hess\ measurements (Aharonian et al. 2007b), shown for comparison. The blue squares are the data binned by observing period (only points with a significance greater than two standard deviations are shown).  The horizontal red lines (the dashed ones represent the statistical error range) are the fit to the \veritas\ yearly data. The black dotted line shows a 3$\%$ Crab Nebula flux for comparison.} \label{VERITASLC}
\end{figure}

These observations resulted in a $\gamma$-ray excess of 489 events, corresponding to a $\gamma$-ray rate of (0.150 $\pm$ 0.014) photons per minute. This corresponds to an average integral flux above 300 GeV of $(23.3 \pm 2.8) \times 10^{-9}$ photons m$^{-2}$ s$^{-1}$, or about 2$\%$ of the Crab NebulaÕs flux. On average, this is a similar flux to that seen by the H.E.S.S. collaboration in 2005 - 2006 (1.8$\%$ of the Crab NebulaÕs flux, above 580 GeV) in 41.8 hours of observation \cite{Aharonian07}. \\
The spectrum shown in Figure \ref{VERITASspec} was fitted with a simple power law, and the resulting normalization (at 1 TeV) and photon index are $(5.54 \pm 0.56_{stat}) \times 10^{-9}$ m$^{-2}$ TeV$^{-1}$ s$^{-1}$ and $2.59 \pm 0.12_{stat}$ respectively, with a $\chi^2$ of $5.8$ with 7 degrees of freedom. The systematic errors on the normalization and index are 20\% and 10\%, respectively. These results are comparable with those seen previously by H.E.S.S., confirming the measured hardness.  The spectral shapes were also derived individually for the first observing period (when the flux was higher, as shown in Fig. \ref{VERITASLC}, and as discussed in further details in Section 4.1), and for the remaining low periods. These are shown as shaded regions in Figure 1. No change is observed in the photon index .

\section{Multi-wavelength observations}

Multi-wavelength observations have been performed, using \swift\ (in X-rays and visible/UV) and \rxte .\\
The \swift\ data set contains sixteen snapshot observations, taken during the 2009-2010 \veritas\ observing season. The \xrt\ data were fitted with an absorbed power law using the absorption calculated by \cite{Sarah} ($N_{H}$ = $1.1 \times 10^{21}$ cm$^{-2}$).  The probability of the \swiftxrt\ flux being constant is 3.8\% and the photon index shows no evidence for variability. Swift Ultraviolet and Optical Telescope (\uvot\ \cite{Poole08}) observations were taken in several different photometric bands (six filters, from visible to UV).  The \uvot\ data were corrected for interstellar extinction (\cite{Seaton79}) and dust absorption (\cite{Schlegel98}). There is still substantial host-galaxy contamination, especially in the B and V bands, which was corrected for using the correction factors derived by \cite{Sarah}. The hard X-ray spectrum (photon index $\Gamma \simeq 1.6$) and UV to X-ray SED suggest that the synchrotron emission peaks above 10 keV (see Fig. \ref{SED}). \\

\onees\ is also in the \swiftbat\ 70-month hard X-ray survey (\cite{BAT}) which includes data from December 2004 to September 2010. This survey contains sources detected
in the \bat\ in the 14 - 195 keV band down to a significance level of 4.8 $\sigma$.  \onees\ is detected
at a flux level of $24.5\pm4.5 \times 10^{-12} $ erg s$^{-1}$ cm$^{-2}$ with a spectral index of $2.16\pm0.28$. The overall SED from the \bat\ is shown in Figure \ref{SED}.\\ 

\onees\ was the target of several \rxte\ (\cite{RXTE}) monitoring campaigns during the \veritas\ observations. This resulted in robust coverage between 2 and 20 keV. There is significant variability in these data, including a large flare before the 2011-2012 \veritas\ observing season.  However, no significant change in the spectral shape is seen during the three years.

\section{Results}
\subsection{Temporal analysis}
Most previous studies using distant blazars to place a lower limit on the IGMF must assume that the measured VHE spectrum (exposure time of the order of tens of hours, accumulated over several months of observations) is a good estimator of the time-averaged spectrum of the source over several years or more. Since the flux from \onees\ and several other VHE blazars had previously been consistent with constant emission, this assumption was made by some authors attempting to limit the strength of the IGMF (\cite{Arlen, Chuck11, Huan11, Neronov10}). However, for variable sources, the multi-year time-averaged differential flux energy density is unknown and difficult to estimate with any reliability. Because of this inherent ambiguity, any lower limit on the IGMF derived using the measured VHE spectrum from variable sources is not robust. The observations presented here show that the constant-flux hypothesis may not be valid for 1ES 0229+200, as shown in Figure \ref{VERITASLC}.\\
 A fit of a constant flux to the \veritas\ yearly points (shown as the solid red line in the figure, $(22.9 \pm 2.80) \times 10^{-9}$ m$^{-2}$ s$^{-1}$) yields a $\chi^2$ of $8.3$ with 2 degrees of freedom (probability of 1.6$\%$). The blue squares in Figure \ref{VERITASLC} show the data divided into individual observing periods (only $>2\sigma$ points are shown; there are 11 additional points which are $<2\sigma$ which have been included in the fit but not in the plot) . A fit of a constant flux to these data results in a $\chi^2$ value of 24.7 with 13 degrees of freedom or a probability of being constant of 2.5$\%$. In conclusion, the evidence for variability in these data is weak but, when considered in the context of a known variable source class, we consider it to be indicative of truly variable emission, at least with a level of confidence which precludes the assumption of a constant flux.\\
 A more significant variability is seen at lower energies, and in particular in X-rays. The \rxte\ data show evidence for longer-term variability, including a large flare preceding the 2011-2012 \veritas\ observing season. A constant flux in the \rxte\ data is excluded at greater than ten standard deviations in all three bands. The hard X-ray data from the \bat\ display an interesting feature. Directly preceding the first season of \veritas\ observations (where the highest VHE fluxes were measured), the \bat\ flux from \onees\ reached a level not previously seen in the lifetime of the \bat\ instrument. The \bat\ flux then dropped to one of the lowest levels seen. The high flux was repeated right at the end of the \veritas\ observing season, where observations at VHE were not possible due to the full moon.\\
 The variability seen in X-rays strengthens the fact that the VERITAS light-curve is not consistent with a constant flux. In fact, in the context of the SSC scenario (see next Section), the same particle population responsible for the X-ray emission (via synchrotron radiation) is associated to the VHE emission (via inverse-Compton scattering). The detection of flaring activity in X-rays implies thus the possibility that a similar variability behavior could indeed be present in the VERITAS data.

\subsection{Spectral analysis}
\begin{figure}[t!]
\includegraphics[width=245pt]{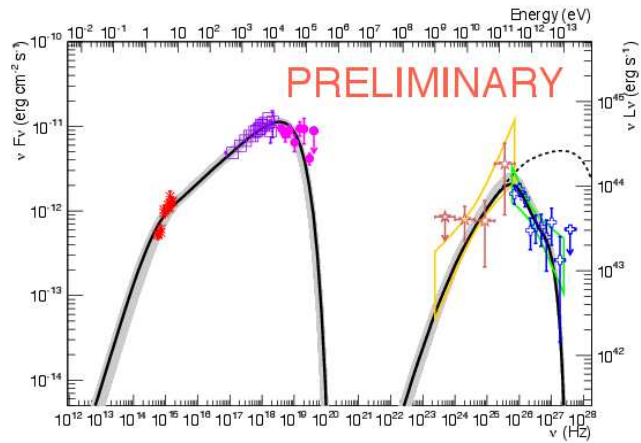}
\caption{The multiwavelength SED of the blazar 1ES 0229+200. From low to high-energies, we report data from \swiftuvot, \swiftxrt, \swiftbat, \fermilat\  and \veritas. The grey lines are all the one-zone SSC models which correctly describe the SED. The solid black curve is the SSC model with the lowest $\chi^2$ value with respect to the data. The dotted black line represents the best-fit SSC model, deabsorbed for the EBL.} \label{SED}
\end{figure}

The broadband SED of \onees\ is shown in Figure \ref{SED}. The \fermilat\ data, as well as the best-fit \lat\ bow-tie are reproduced from \cite{Vovk12}, where they reported a detection at the level of seven standard deviations in almost three years of observations. Note that the \swiftbat, \fermilat, and \veritas\ data are average spectra, while the \swiftxrt\ and \uvot\ data are from the initial \veritas\ observing season in 2009-2010. The availability of \swift, \fermi\ and \veritas\ data allows both the low-energy and high-energy peaks to be constrained. The \swiftxrt\ spectrum is especially hard, indicating that the synchrotron peak is located above the XRT energy band, but the additional information from the BAT suggests that the peak is located between the two bands at $E \simeq 10$ keV.\\
The SED is modeled using the one-zone synchrotron-self-Compton (SSC) code of \cite{Katar01}, taking into account EBL attenuation based upon the calculation of \cite{Franceschini08}. The SSC parameter space is constrained by the numerical algorithm described in \cite{Cerruti13}. The model assumes a spherical emission volume of radius $R$ moving towards the observer with Doppler factor $\delta$, and filled with a tangled, homogeneous magnetic field $B$ and a non-thermal population of electrons and/or positrons $N_e(\gamma_e)$. The particle distribution is parameterized by a single power-law function (with index $\alpha$), between minimum and maximum Lorentz factors $\gamma_{min}$ and $\gamma_{Max}$, respectively. Note that the modeling presented here ignores any possible contribution to the SED from photons reprocessed by the IGMF.\\
Thus, the model has seven free parameters: $\delta$, $B$ and $R$, for the emitting region, and $\alpha$, $\gamma_{min;Max}$ and $K$, for the particle distribution, where $K$ is defined as the electron number density at $\Gamma_{Max}$. In order to determine the set of solutions which correctly describe the SED, we used seven physical observables: the synchrotron peak frequency and flux, the X-ray spectral index, the \fermi\ and \veritas\ fluxes at their respective decorrelation energies, and the \fermi\ and \veritas\ spectral indices. In addition, the break observed in the optical/UV data after correction for the host-galaxy contamination provides a separate constraint which significantly narrows the range of solutions.\\
  
   \begin{table}[t!]
\begin{center}
\caption{Preliminary. Best-fit model parameters for \onees. For each parameter we report the minimum and maximum value obtained. Note that the model parameters are correlated. }
\label{results}
\begin{tabular}{c|c}
\hline
\hline 
$\delta$ & $56.4-100$\\
$B$ & $7.5\times10^{-4}-2.6\times10^{-3}$ G\\
$R$ & $5.8\times10^{15}-4.9\times10^{16}$ cm\\
$\gamma_{min}$ & $2.5\times10^4-4.5\times10^4$ \\
$\gamma_{Max}$ & $3.3\times10^6-8.8\times10^6$\\
$K$ & $9.8\times10^{-13}-7.7\times10^{-11}$ cm$^{-3}$ \\
$\alpha$ & 2.3\\
\hline
\hline
\end{tabular}
\end{center}
\end{table}

The best-fit parameter values are given in Table \ref{results}.  Previous SSC modeling attempt of the SED of \onees\ have been made by \cite{Tavecchio09} and \cite{Sarah}. The main difference with respect to these previous models comes from the value of the synchrotron peak frequency, which, in our SED, is located between the XRT and the BAT energy bands, at $\nu_{sync-peak} = 3\times 10^{18}$ Hz, an order of magnitude less than that reported by \cite{Sarah} ($3.5\times 10^{19}$ Hz), but more in line with what \cite{Tavecchio09} found ($9.1\times 10^{18}$ Hz). This difference arises mainly from the higher statistics in the 70-month \bat\ spectrum compared with the 58-month spectrum previously used. Another difference compared to the previous modeling attempts of \onees\ in an SSC scenario is that neither \cite{Tavecchio09} or \cite{Sarah} had the information from the Fermi-LAT detection.\\
The first important result of the current modeling is that the minimum value of the Doppler factor required to fit the SED of \onees\ is $\delta \geq 56$. This value is higher than the ones commonly assumed in SSC modeling of HBLs (see, for example, \cite{Mrk501, Mrk421, 2155}) but in agreement with the one adopted by \cite{Tavecchio09}($\delta=50$), while \cite{Sarah} adopted $\delta=40$.\\
In our modeling, we found that a parameterization of the electron distribution by a single power-law function provides a good description of the SED. However, a break in the spectrum is expected in the presence of synchrotron cooling, but it is possible that the break energy is above the value of $\gamma_{Max}$. Following the study presented in \cite{Cerruti13}, we evaluate that this is possible only if the injected particles are escaping the emitting region with a speed $v \le c/50$. Another interesting aspect is the energy budget of the emitting region. For all the solutions, we compute the values of the magnetic and the particle energy density. The equipartition factor $u_e/u_B$ is comprised between $2\times10^4$ and $10^5$, implying an emitting region significantly out of equipartiton. \\
The final relevant point is that the value of $\gamma_{min}$, constrained between $2.5\times10^4$ and $4.5\times 10^4$, is unusually high compared to standard SSC modeling of blazars. The fact that the modeling of hard- VHE-spectrum HBLs requires such a high value of $\gamma_{min}$ has been previously noted by \cite{Kata06}, who claimed that $\gamma_{min} \geq 10^4$ can be a characteristic of this kind of source. \\

\section{Conclusions}
VERITAS performed a long-term observation of the VHE HBL \onees\ from 2010 - 2012 for a total time of 54.3 hours, providing the most detailed VHE SED of this blazar to date. The overall average integral flux during this time was $(23.3 \pm 2.8) \times 10^{-9}$ photons m$^{-2}$ s$^{-1}$ (E $>$ 300 GeV) and the spectrum is well described by a power law with photon index $\Gamma=2.59\pm0.12_{stat}$. A fit of a constant flux to the \veritas\ yearly points yields a probability of 1.6\% ($\chi^2$=8.3 for 2 degrees of freedom). No significant change in spectral shape is seen.\\
The VHE observations were supported by several multiwavelength data sets ranging over many orders of magnitude in energy from optical to GeV. This allowed for detailed SED modeling based on the code of \cite{Katar01}.  The Doppler factor of $\delta \geq 56$ is similar to that adopted by \cite{Tavecchio09}.  The observations of 1ES 0229+200 presented here are part of the VERITAS long-term blazar observing program. This program was developed to build up a database of SEDs from a variety of blazars. Under these auspices, we have produced the most detailed SED measurement of this hard-spectrum distant blazar to date, and we have discovered evidence for variability at VHE. Regular VERITAS observations of 1ES 0229+200 are continuing which will be used to further characterize the SED and the nature of the underlying variability.

\vspace*{0.5cm}
\footnotesize{{\bf Acknowledgment:}{
This research is supported by grants from the U.S. Department of Energy Office of Science, 
the U.S. National Science Foundation and the Smithsonian Institution, by NSERC in Canada, by 
Science Foundation Ireland (SFI 10/RFP/AST2748) and by STFC in the U.K. We acknowledge the excellent
work of the technical support staff at the Fred Lawrence Whipple Observatory and at the collaborating 
institutions in the construction and operation of the instrument. This research has made use of the NASA/IPAC Extragalactic Database (NED) which is operated by the Jet Propulsion Laboratory, California Institute of Technology, under contract with the National Aeronautics and Space Administration.}
}

\end{document}